# Real Differences between OT and CRDT in Building Co-Editing Systems and Real World Applications

DAVID SUN, Codox Inc., United States
CHENGZHENG SUN, Nanyang Technological University, Singapore
AGUSTINA NG, Nanyang Technological University, Singapore
WEIWEI CAI, Nanyang Technological University, Singapore

OT (Operational Transformation) was invented for supporting real-time co-editors in the late 1980s and has evolved to become core techniques widely used in today's working co-editors and adopted in industrial products. CRDT (Commutative Replicated Data Type) for co-editors was first proposed around 2006, under the name of WOOT (WithOut Operational Transformation). Follow-up CRDT variations are commonly labeled as "*post*-OT" techniques capable of making concurrent operations natively commutative in co-editors. On top of that, CRDT solutions have made broad claims of superiority over OT solutions, and often portrayed OT as an incorrect and inefficient technique. Over one decade later, however, CRDT is rarely found in working co-editors; OT remains the choice for building the vast majority of today's co-editors. Contradictions between the reality and CRDT's purported advantages have been the source of much confusion and debate in co-editing researcher and developer communities. To seek truth from facts, we set out to conduct a comprehensive and critical review on representative OT and CRDT solutions and working co-editors based on them. From this work, we have made important discoveries about OT and CRDT, and revealed facts and evidences that refute CRDT claims over OT on all accounts. These discoveries help explain the underlying reasons for the choice between OT and CRDT in the real world. We report these results in a series of three articles.

In prior two papers, we have reported our discovery that CRDT *is not* natively commutative for concurrent operations in co-editors as commonly claimed, but *is* like OT in following a general transformation approach to consistency maintenance, albeit *indirectly*. Furthermore, we have reported the differences between OT and CRDT in realizing the same general transformation approach in co-editors, and the consequential differences in correctness and complexity. In this article (the third in the series), we examine the role of building working co-editors in shaping OT and CRDT research and solutions, and consequential differences in the choice between OT and CRDT in real world co-editors and industry products. In particular, we review the evolution of co-editors from research vehicles to real world applications, and discuss representative OT-based co-editors and alternative approaches in industry products and open source projects. Moreover, we evaluate CRDT-based co-editors in relation to published CRDT solutions, and clarify some myths surrounding "peer-to-peer" co-editing. We hope the discoveries from this work help clear up common myths and confusions surrounding OT and CRDT, and accelerate progress in co-editing technology for real world applications.

CCS Concepts: • **Information Systems → Group and Organization Interfaces**; Synchronous Interaction, Theory and Model.

## KEYWORDS

Operational Transformation (OT); Commutative Replicated Data Type (CRDT); Concurrency Control; Consistency Maintenance; Real-Time Co-Editing; Cloud/Internet/Distributed Computing; Computer Supported Cooperative Work (CSCW) and Social Computing.

## 1 INTRODUCTION

Real-time co-editors allow multiple geographically dispersed people to edit shared documents at the same time and see each other's updates instantly [1,6,14,15,16,17,39,44,55,56,61,73,79]. One major challenge in building such systems is consistency maintenance of documents in the face of concurrent editing, under high communication latency environments like the Internet, and without imposing interaction restrictions on human users [14,55,56].

**Corresponding author**: Chengzheng Sun, School of Computer Science and Technology, Nanyang Technological University, Singapore. Email: CZSun@ntu.edu.sg; URL: https://www.ntu.edu.sg/home/CZSun

Operational Transformation (OT) was invented to address this challenge [14,55,62,73] in the late 1980s. OT introduced a framework of transformation algorithms and functions to ensure consistency in the presence of concurrent user activities. The OT framework is grounded in established distributed computing theories and concepts, principally in *concurrency* and *context* theories [25,55,67,68,84,85]. Since its inception, the scope of OT research has evolved from the initial focus on consistency maintenance to include a range of key collaboration-enabling capabilities, including *group undo* [39,45,58,59,67,68], and *workspace awareness* [1,20,61]. In the past decade, a main impetus to OT research has been to move beyond plain-text co-editing [6,14, 21,39,44,55,56,59,62,71,72,78], and support real-time collaboration in rich-text co-editing in word processors [61,66,69,83], HTML/XML Web document co-editing [11], spreadsheet co-editing [70], 3D model co-editing in digital media design tools [1,2], and file synchronization in cloud storage systems [3]. OT-based co-editors have also evolved from supporting people to use the same editor in one session (*homogeneous* co-editing), to allowing the use of different editors in the same session (*heterogeneous* co-editing) [9]. Recent years have seen OT being widely adopted in industry products as the core technique for consistency maintenance, ranging from battle-tested online collaborative rich-text editors like Google Docs[1][12], to emerging start-up products, such as Codox Apps[2].

A variety of alternative techniques for consistency maintenance in co-editors had also been explored in the past decades [15,17,19,42,43,73]. One notable class of techniques is CRDT[3] (Commutative Replicated Data Type) for co-editors [4,5,8,26,33,38,40,41,42,46,48,49,80,81,82]. The first CRDT solution for plain-text co-editing appeared around 2006 [41,42], under the name of WOOT (WithOut Operational Transformation). One motivation behind WOOT was to solve the *FT* (*False Tie*) puzzle in OT [54,56,75], using a radically different approach from OT. Since then, numerous WOOT revisions (e.g. WOOTO [81], WOOTH [4]) and alternative CRDT solutions (e.g. RGA [46], Logoot [80,82], LogootSplit [5]) have appeared in literature. In CRDT literature, CRDT has often been labeled as a "*post*-OT" technique that makes concurrent operations *natively* commutative, and does the job "*without operational transformation*" [41,42], and even "*without concurrency control*" [26]. On top of that, CRDT solutions have made broad claims of superiority over OT solutions, in terms of *correctness,* time and space *complexity*, *simplicity*, etc.

After over one decade, however, CRDT solutions are rarely found in working co-editors or industry co-editing products, and OT solutions remain the choice for building the vast majority of co-editors. The contradictions between the reality and CRDT's purported advantages over OT have been the source of much confusion and debate in co-editing research and developer communities. Have the majority of co-editors been unfortunate in choosing the faulty and inferior OT, or such CRDT claims are false? What are the real differences between OT and CRDT for co-editors? What are the key factors that may have affected the choice between OT and CRDT for co-editors in the real world? We believe that a thorough examination of these questions is relevant not only to researchers exploring the frontiers of collaboration-enabling technologies and systems, but also to practitioners seeking viable techniques to build real world collaboration tools and applications.

To seek answers to above questions and beyond, we set out to conduct a comprehensive and critical review of representative OT and CRDT solutions and working co-editors based on them, which are available in publications or from publicly accessible open-source project repositories. In this work, we explored *what, how,* and *why* OT and CRDT solutions are different and the consequences of their differences from both an algorithmic angle and a system perspective. From this exploration, we made a number of discoveries, some of which are rather surprising. One such discovery is that CRDT is not natively commutative for concurrent operations for co-editors as commonly claimed, but actually is the same as OT in following a general transformation approach to achieving consistency in co-editors. This study has also examined major CRDT claims over OT and provided facts and evidences that refute those claims on all accounts.

---

[1] https://www.google.com/docs/about/

[2] https://www.codox.io

[3] In literature, CRDT can refer to a number of different data types [48, 49]. In this paper, we focus *exclusively* on CRDT solutions *for text co-editors*, which we abbreviate as "CRDT" in the rest of the paper, though occasionally we use "CRDT for co-editors" for emphasizing this point and avoiding misinterpretation.

We have focused this study on OT and CRDT solutions to *consistency maintenance* in *real-time* co-editing, as it is the foundation for other co-editing capabilities, like group *undo* and issues related to *non-real-time* co-editing, which we plan to cover in future work. Also, we focus our discussions on basic issues in plain-text co-editing as the capability of supporting plain-text co-editing is the foundation for support co-editing in more advanced domains (e.g. rich text co-editing), and the vast majority of published CRDTs are confined in this domain. We know of no existing work that has made similar attempts.

The topics and bulk of outcomes from this study are comprehensive, complex and diverse, and have different accessibilities to readers with different interests and backgrounds. To cope with the complexity and diversity and take into account of feedback to a prior version of our report on this work (see footnote 4 in [74] and [75]), we have organized the materials into three parts and presented them in a series of three related but self-contained articles.

In [74] (the first paper of this series), we have examined the real differences between OT and CRDT under a general transformation framework for consistency maintenance in co-editors. In particular, we review the basic ideas of OT and CRDT and present a general transformation framework for consistency maintenance in co-editors. Furthermore, we reveal that CRDT *is not* natively commutative for concurrent operations in co-editors as often claimed, but *is* like OT in following the general transformation approach, albeit *indirectly.* Uncovering the hidden transformation nature and demystifying the commutativity property of CRDT provides much-needed clarity about what CRDT really *is* and *is not* to co-editing, which in turn brings out the real differences between OT and CRDT for co-editors.

Built on the outcomes from [74], we have further examined the real differences between OT and CRDT in correctness and complexity for consistency maintenance in co-editors in [75] (the second paper of this series). We dissect representative OT and CRDT solutions, and explore how different basic approaches to realizing the same transformation – the *concurrency-centric* and *direct* transformation approach taken by OT versus the *content-centric* and *indirect* transformation approach taken by CRDT – had resulted in different technical challenges, consequential correctness, and complexity issues. Moreover, we reveal hidden complexity and algorithmic flaws within representative CRDT solutions, and discuss common myths and facts related to correctness, time and space complexity, and simplicity of OT and CRDT solutions.

In this paper (the last one in the series), we examine the real differences between OT and CRDT in building co-editing systems and supporting real world applications. We examine the role of building co-editing systems in shaping OT and CRDT research and solutions, respectively. Also, we discuss representative OT-based co-editors and alternative system approaches in industry products and open source projects. Moreover, we evaluate CRDT-based co-editors in relation to published CRDT solutions, and clarify myths surrounding "peer-to-peer" co-editing.

## 2 BUILDING CO-EDITORS BASED ON OT
### 2.1 The Role of Building Co-Editors in OT Research

While theoretic work around OT algorithms has been fundamental in OT research, the practice of building OT-based working co-editors has played a crucial role in driving and shaping OT research for over two decades [1,2,14,37,39,44,56,59,60,61,63,65,66,69,83]. Designing and implementing co-editors have enabled researchers to identify topics that are truly relevant and important to co-editing, uncover intricate technical issues that would otherwise go unnoticed by pure theoretical study, motivate innovative technical solutions, experimentally validate solutions from system perspectives, and gain critical insights for deriving general principles and theories, which in turn inspire and guide experimental exploration.

*2.1.1 Building and Using Co-Editors as Research Vehicles*
The very first OT research publication detailed the design and implementation of a plain-text co-editor GROVE [14]. A succession of working co-editors, including DistEdit [39], Jupiter [37],

JOINT EMACS [44], and REDUCE [54,56,59], were built by researchers to investigate both system and theoretical issues in co-editing. These experimental efforts revealed critical insights into the dOPT puzzle (a flaw in the first OT control algorithm named dOPT [14,55]), and eventually led to the resolution of this puzzle and the establishment of the theoretic foundation for OT − a comprehensive set of transformation conditions and properties, such as context-based conditions and transformation properties [39,44,55,56,62].

Early OT-based co-editors mainly served as research vehicles to investigate novel consistency maintenance techniques for co-editing plain-text documents, but placed little emphasis on the relevance to supporting real world applications that users may use daily for content creation [18].

*2.1.2 Applying OT to Real World Co-Editors and the TA Approach*

It was around the year 2000 when researchers began to investigate the possibility of extending OT from supporting plain-text documents to off-the-shelf productivity suites with complex document formats and comprehensive functionalities. The Transparent Adaptation (TA) approach is one representative work along the line of exploration [61,66,83].

The goal of the TA approach was to extend the basic OT to support complex applications beyond text editing and to convert single-user editors into co-editors, without changing the source code of the original applications. The concurrency-centric nature of OT ensures the core control algorithms to be generic and allows transformation functions to be extensible to new application domains. Leveraged on this OT property, the TA approach is able to handle a myriad of complex data objects and user interactions found in modern productivity applications for real-time co-editing. A set of diverse productivity applications were examined and successfully converted into co-editors, including Microsoft Word (CoWord [61,66,69,83]), PowerPoint (CoPowerPoint [61,66]), and Autodesk Maya (CoMaya [1,2]).

A TA-based co-editor consists of three architectural components [61], as shown in Fig. 1:

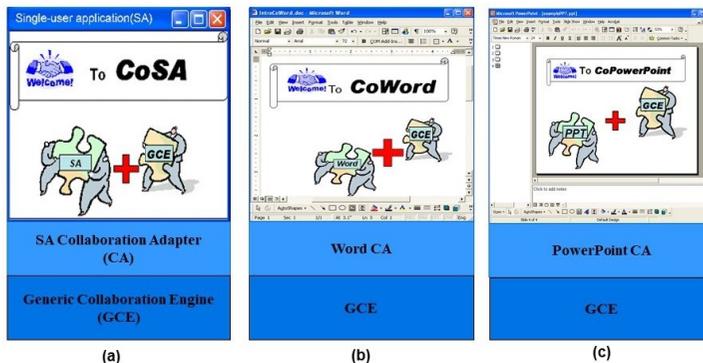

Fig. 1. (a) A TA-based architecture for co-editors. (b). CoWord architecture. (c) CoPowerPoint architecture.

1. Generic Collaboration Engine (GCE), which provides generic transformation capabilities (independent of specific OT control algorithms or transformation function).
2. Collaboration Adaptor (CA), which bridges the GCE with a single-user application, effectively extending basic transformation functionalities to the single-user application.
3. A Single-user Application (SA), which provides conventional editing features and functionalities to users and suitable API (Application Programming Interface) to the CA.

One key insight embedded in the TA architecture is: multi-user collaboration capabilities and single-user conventional editing functionalities are *orthogonal* and can be encapsulated in different components of a co-editing system [61]. The TA architecture provides a general framework for separating collaboration capabilities (powered by suitable techniques, e.g. OT) from conventional editing functionalities (encapsulated in existing or newly designed single-user editors) and integrating these two in a co-editor.

With the TA approach and a reusable GCE, the task of building a new co-editor is reduced to building a new CA for bridging a single-user editor with the GCE, without reinventing a new editor (if an existing editor is used) or re-implementing an OT-based collaboration engine. Potential challenges and opportunities for innovation in building TA-based co-editors lie inside the CA component. The TA approach has since had important impact on follow-up co-editing research and system design in the real world (see Section 2.2).

*2.1.3 Extending OT for Supporting Complex Data and Operation Models*

Building TA-based CoWord, CoPowerPoint and CoMaya systems was the main stimuli behind a number of key leaps in co-editing technology innovation. A major innovation was the extension of the basic OT data model from a single linear addressing space (for text editors) to a tree of multiple linear addressing domains ─ XOTDM (eXtended OT data model), which can be used to model complex documents in rich-text editors, word processors, and digital design tools [1,2,61].

Complementing the OT data model extension was the extension of the OT operation model from supporting primitive *insert* and *delete* operations to arbitrary complex application operations. For an editing system capable of supporting $N$ application operations, one school of thought is to design $N \times N$ transformation functions, one function for transforming each pair of application operations. We label this approach as AOT (Application Operation Transformation). This AOT approach can work well for editors supporting a relatively small number of operations, e.g. a plain-text editor with two operations *insert* and *delete*, where four pair-wise transformation functions will be adequate [11,14,56,62,72]. However, designing direct application transformation functions is challenging for editors when the number of operations grow and when the semantics of those operations become complex (such as those in Word, PowerPoint and Maya). In addition, because application operations are by nature application-specific, their transformation functions are not reusable across different applications, which means significant redesign and validation efforts are needed for supporting a new co-editing application.

An alternative approach to supporting arbitrary application operations, pioneered in the TA-based CoWord, consists of two parts: (1) a collection of transformation functions for a *small* number of *primitive* operations, e.g. *insert*, *delete*, and *update* (for capturing a range of general editing functionalities ─ an important extension to the basic OT operation model); and (2) a collection of *adaptation* schemes that translate arbitrary application operations to/from primitive operations. We label this transformation approach as POT+COA (Primitive Operation Transformation *plus* Complex Operation Adaptation). The merit of this approach is that transformation functions for primitive operations are relatively easy to design and reusable across different applications; the challenge with this approach lies in designing the adaptation schemes between complex application operations and primitive operations. Experiences in applying this approach across a spectrum of TA-based co-editing applications have shown that devising adaptation schemes between complex application operations and primitive operations are relatively easier and more robust than designing direct AOT functions among complex application operations. For detailed discussion on issues related to adaptation schemes in TA-based co-editors, the reader is referred to [1,61,66,83].

*2.1.4 Interactions between Academia and Industry in OT Research*

The first OT was originated from an industry research lab (Microelectronics and Computer Consortium (MCC)) in 1989 [14]. A few years later, several research groups independently discovered and resolved a fundamental algorithmic flaw (i.e. the well-known dOPT puzzle) [55], which revived interests on OT and helped establish OT, as well as co-editing technical research, as a niche system area in CSCW.

Follow-up OT research activities were mostly conducted in universities and research institutes. Significant advancements were made along the lines of OT functionality extensions (e.g. from consistency maintenance to group undo, operation compression, etc.), algorithm design, puzzle detection and resolution, correctness verification, and complexity analysis [75]. Those academic OT works nevertheless attracted little interest from industry. OT work was mostly confined within the academic world until researchers started to extend OT to real world applications, and successfully built systems, such as CoWord, CoPowerPoint, and CoMaya. These non-toy systems demonstrated OT relevance to the real world, which helped bridge and reconnect academia and industry [34].

Subsequently, academia researchers and industry practitioners have interacted in a number of forms, including technical talks at industry labs (e.g. [60,63]), public demos of working prototypes (e.g. [60,63,65]), and tutorials on co-editing technologies (e.g. [64]), which were attended by people from both universities and industry, at ACM CSCW conferences. The co-editing research community and members of industry have also jointly organized a series of co-editing workshops (e.g. [36]), to share experiences in building and using co-editing systems.

## 2.2 OT-based Industry Co-Editing Products and Techniques

### 2.2.1 OT Adoption in Major Industry Products

In 2009, Google announced adoption of OT as the core technology for supporting its real-time collaboration features in Google Wave [77,79]. Since then, there has been an explosion of interests in OT and co-editing from computer industry and open source software communities.

In 2010, OT was adopted in Google Docs [12] ─ a Web-based real-time collaborative office suite. Google Wave OT algorithms and protocols were handled over to Apache Software Foundation under the name Apache Wave[4]. It had strong influences on a number of Web-based open-source OT software projects. One representative open-source OT project is ShareJS[5]. Other notable OT-based co-editors include SubEthaEdit[6], CKEditor[7], Etherpad[8], to name a few.

In later years, cloud storage companies started to extend their storage and file synchronization services to offering new Web-based co-editing services on the files in their storage. These companies built their own OT-based rich-text co-editors, such as Dropbox Paper[9] (2017), and Box Notes[10] (2017). More recently, Tencent also integrated the OT-based rich-text co-editing capability into its cloud-based TAPD (Tencent Agile Product Development) (2018)[11].

Most commercial collaborative editing products, such as Google Docs, etc., have been designed and implemented for collaborative editing from scratch: from document markups and formats, editing operations, interface features, to application-specific transformation functions for supporting co-editing. However, building feature-rich and robust co-editors from scratch involves major investment of engineering resources, which is unaffordable to many. Furthermore, these co-editors are often vertically integrated with content storage solutions, which lock users in specific cloud services. For example, users need to move and store their contents to specific repositories, such Google Drive, Dropbox, and Box, in order to use these co-editors.

### 2.2.2 Alternative Approaches to Building OT-based Co-Editors on the Web

In contrast to Google Docs and other Web-based co-editors built from scratch, Codox Apps were built without reinventing new editors for conventional editing functionalities or re-implementing core OT solutions for each editor in supporting real-time collaboration capabilities.

Codox supports real-time co-editing directly in existing Web-based applications, such as Gmail[12], Evernote[13], WordPress[14], Zendesk[15], Wikipedia[16], TinyMCE[17], Froala[18], Quill[19], ClickUp[20], and Slate[21], and retain functionalities and the "look-and-feel" of original applications. In each Codox Apps, an application-specific adaptor is injected into the single-user editing environment, which bridges an existing editor and an OT-powered reusable Codox engine, as shown in Fig. 2. Codox has also provided public APIs (of the generic Codox engine, application-specific adaptors, and real-time cloud services) for other people to convert their favorite single-user editors into co-editors in cloud-based collaborative environments.

---

[4] https://en.wikipedia.org/wiki/Apache_Wave
[5] https://github.com/share
[6] https://www.codingmonkeys.de/subethaedit/
[7] https://ckeditor.com/collaborative-editing/
[8] http://etherpad.org/
[9] https://paper.dropbox.com/
[10] https://www.box.com/notes
[11] https://www.tapd.cn/
[12] https://chrome.google.com/webstore/detail/wave-for-gmail-and-everno/dggkchdpgkalbmhnlmgmiafjacofjghb?utm_source=inline-install-disabled
[13] https://evernote.com/
[14] https://wordpress.org/plugins/wave-for-wp/
[15] https://www.zendesk.com/apps/support/wave/
[16] https://www.wikipedia.org/
[17] https://docs.codox.io/integrations/tinymce
[18] https://froala.com/wysiwyg-editor/examples/codox-real-time-editing/
[19] https://codepen.io/dnus/pen/OojaeN
[20] https://docs.clickup.com/en/articles/1427936-collaborative-editing-collaboration-detection/
[21] https://codepen.io/dnus/pen/yZdrwX

Many OT-based Web co-editors have adopted the AOT approach in designing transformation function, such as those transformation functions designed in Google Wave [79] and Docs [12], CKEditor[22], etc. In contrast, the Codox Apps have taken the POT+COA approach to supporting complex application operations. Also, Codox has taken (and extended) the XOTDM data model [61] as the basis to support collaborative Web applications using hierarchical data models (e.g. JSON[23]). See Section 2.1.3 for discussion on XOTDM, and pros and cons of AOT and POT+COA.

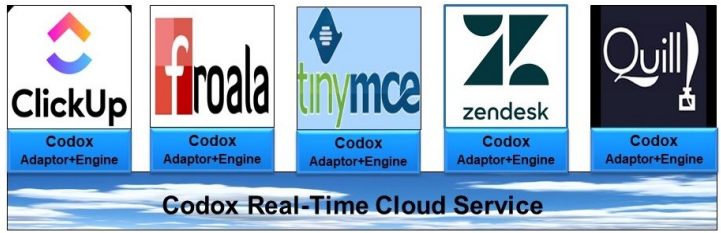

Fig. 2. Codox App architecture and example apps.

Unlike many OT-based Web co-editors which use central OT servers (i.e. server-based OT solutions), Codox Apps do not need any central OT server (i.e. server-less or distributed OT solutions). We give more elaboration on server-based versus sever-less OT solutions and system architectures of OT-based co-editors in the next Section 2.3.

Despite various technical differences among Web-based co-editors, one common system design approach have emerged, i.e. to separate and encapsulate collaboration capabilities and conventional editing functionalities in different functional components, and provide ways to integrate these components together in a complete co-editing system, which is like the TA approach (see Section 2.1.2). For example, the Codox approach is clearly a Web-based version of the TA approach. ShareJS and Convergence Labs[24] support co-editor design by providing packages of OT-powered co-editing capabilities, which can be integrated with separately designed single-user editing applications. QuillJS[25] and SlateJS[26] provide different frameworks for separately designing single-user rich text editors and multi-user collaboration capabilities (based on OT and/or other techniques), and offer APIs to bridge the two parts together to create co-editors. These alternative approaches are complementary and have shown to be promising to realize the dream of making real-time co-editors commonplace on the Web in the near future.

Last decade has witnessed significant efforts from industry and open source communities in applying OT to numerous real world applications, with large-scale deployment. The theory and practice of OT is now driven by a vibrant community of academic researchers and industry practitioners. A wealth of valuable experiences have been accumulated from building and using co-editing applications, which calls for comprehensive technical reviews of real world system approaches, issues and experiences in the future.

## 2.3 Architectures and Communication Topologies of OT-based Co-Editors

OT-based co-editors have been built with a variety of system architectures and communication topologies, e.g. a *star-like* communication topology among co-editing clients, with a central server for broadcasting message and with *optional* transformation capabilities, or a *fully connected* topology without a central server for message broadcasting or transformation.

---

[22] https://ckeditor.com/blog/Lessons-learned-from-creating-a-rich-text-editor-with-real-time-collaboration/

[23] https://en.wikipedia.org/wiki/JSON

[24] https://convergence.io/

[25] https://github.com/quilljs/delta

[26] https://github.com/aha-app/collaborative-demo

The choice of a particular system architecture and communication topology depends on multiple technical factors in co-editing, such as where to store and access shared documents for co-editing, how to manage co-editing sessions (e.g. starting, joining and quitting sessions), how to propagate co-editing messages, and what kind of OT solution is used in the co-editor. In this context, whether the OT solution is a *server-based* or *distributed* solution is particularly relevant, which are discussed below.

*2.3.1 Server-based versus Distributed OT Solutions*
There are two general classes of OT solutions: *server-based* OT solutions, and *distributed* (or *server-less*) OT solutions. A server-based OT solution requires an OT-based server and runs different OT algorithms at the server and client sites in a co-editing session, as shown in Fig. 3(a). Example server-based OT solutions include those in Jupiter [37], Google Wave or Docs [12,35,79], NICE [50], CKEditor, ShareJS, QuillJS, and SlateJS.

A distributed OT solution does not require an OT-based server and the same OT algorithm run at all co-editing sites, as shown in Fig. 3 (b) and (c). Representative distributed OT solutions include adOPTed [44], GOT [54,56], GOTO [55], TIBOT [27,85], COT [67,68], POT[85], and SOCT2 [52]. SOCT3/4 [78] is special: it uses a server to issue continuous total ordering numbers, but this server does not perform any transformation.

*2.3.2 OT-based Co-Editing System Architectures and Communication Topologies*
While server-based OT solutions require a server-based co-editing system architecture (for running the server part of OT solutions), distributed OT solutions do not dictate the choice of specific architectures or communication topologies for co-editors based on them.

If a co-editor has adopted a distributed OT solution, it has various communication topology options, e.g. to use a message server with a star-like communication topology for broadcasting messages and other purposes, illustrated in Fig. 3 (b); or a fully connected communication topology for message broadcasting without involving a server, illustrated in Fig. 3 (c). The mere existence of a server in an OT-based co-editing system does not imply the server is necessarily required by the OT solution. Example co-editors based on distributed OT solutions include JOINT EMACS [44], REDUCE [54,56,59], CoWord [61], CoMaya [1,2], and numerous Codox Apps.

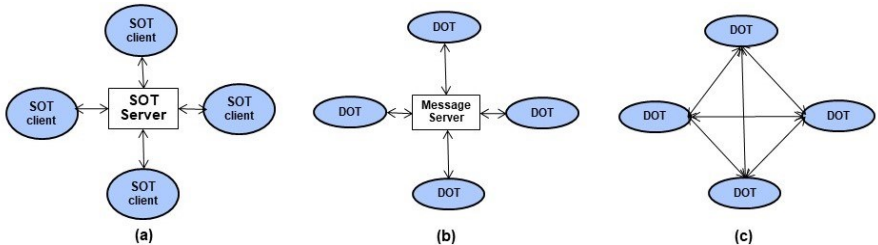

Fig. 3. Varieties of OT architectures and communication topologies. (a). Server-based OT (SOT) solution using a server for OT and communication (e.g. OT in Google Wave or Docs). (b). Distributed OT (DOT) solution using a server for communication (e.g. OT in CoWord); (c) Distributed OT (DOT) solution with full-connection for communication (e.g. OT in REDUCE).

*2.3.3 Misconceptions about OT System Architectures*
In co-editing literature, there is one common misconception about OT, which says that *a central server is required for OT to work*. This is clearly incorrect as there exist numerous distributed OT solutions that require no central server to do its work.

Furthermore, there are two common misconceptions surrounding distributed OT solutions as well: (1) one is that distributed OT solutions require transformation functions to preserve CP2 [75]; (2) the other is that distributed OT solutions have to use vector-based timestamping. To refute these misconceptions, it suffices to give a few counter-examples, e.g. POT [85] and TIBOT/TOBOT2.0 [27,85] are all distributed (sever-less) OT solutions that do not require transformation functions to preserve CP2, and are based on scalar timestamping. For more detailed discussions on how to achieve CP2-correctness in OT, the readers are referred to [75] (in

Section 3.3. "Solving the CP2 Issue and the FT Puzzle under the OT Approach") and [85] (on general conditions and concrete strategies taken by seven OT solutions capable of avoiding CP2).

In reality, existing co-editors have commonly used servers to support some aspects of co-editing, e.g. shared document storage, session management, co-editing message broadcasting, and *optionally* for running server-side OT algorithms if a server-based OT solution is used. To our best knowledge, there has been no working co-editor based on OT (or other techniques) that does not use a central server, but the server is often used for supporting aspects of co-editing, *not* necessarily due to the need of OT.

## 3 BUILDING CO-EDITORS BASED ON CRDT

### 3.1 The Role of Building Co-Editors in CRDT Research

The first CRDT (WOOT) started as an effort to solve the FT puzzle and as an alternative to OT for supporting plain-text co-editing around 2006, which was several years after OT research had started to support real world applications and built rich-text co-editors like CoWord and CoPowerPoint (Section 2.1). Since the start, CRDT research has adopted predominantly theoretic approaches to identifying research issues, designing and verifying solutions (e.g. using theorem provers, model checkers, etc.) [7,22,23,41,42,43], but rarely implemented CRDT solutions in working co-editors for validation. These approaches have had profound impact in shaping CRDT research and its adoption in the real world.

### 3.2 CRDT-based Co-Editors Built by Practitioners

We found no working CRDT-based co-editors built by CRDT researchers or academic literature documenting system issues and experiences in using proposed CRDT solutions for building working co-editors. We did, however, find a dozen of CRDT-based co-editor projects hosted on GitHub, which were created by practitioners who were interested in learning whether and how CRDT solutions actually worked when applied to realistic editing environments.

*3.2.1 Observations on CRDT-based Plain Text Co-Editors*

Most of those implementations are at rudimentary stages of development, but we found two relatively stable plain-text co-editing prototypes: Teletype[27], which is based on WOOT (with tombstones) [8,41,42], and Alchemy Book[28], which is based on Logoot (without tombstones) [80,82]. Based on our experimentation with live demos and review of available documentation and source code of those prototypes hosted at GitHub, we made the following observations.

First, *CRDT-based co-editors were mostly built by combining a CRDT solution with an existing text editor*. For example, Teletype was built by integrating WOOT with a desktop text editor named Atom[29]; Alchemy Book was built by integrating Logoot with a Web-based text editor CodeMirror[30]. The implementations of Teletype and Alchemy Book show similarities with the TA approach [61]. Such similarities are unsurprising since TA is based on the general transformation but independent of specific transformation solutions, and CRDT has been shown to be an instance of the general transformation [74]. Those CRDT-based co-editors also show that CRDT object sequences and identifier-based operations are *not* native to real editors.

Second, *CRDT-based co-editor implementations revealed key steps missed in CRDT publications.* We examined how co-editing is supported *end-to-end* under CRDT-based co-editors, i.e. from the point when a user generates an operation from a local document, all the way to the point when this operation is applied to a remote document seen by another user. Both Teletype and Alchemy Book, as well as other CRDT-based co-editor implementations, unmistakably convert user-generated *position-based* operations into *identifier-based* operations at local sites (which is *obscured* in Logoot [80,82]), and convert remote *identifier-based* operations, after applying them in internal object sequences, to *position-based* operations at remote sites (which is *ignored* in WOOT [8,41,42] and RGA [46]). This observation confirms our illustration in Fig. 1-(c) in [74], and our description of CRDT under the general transformation framework in [74]. These missing steps are *not* mere implementation details, but *crucial* steps for CRDT solutions to achieve consistency maintenance in co-editors. Unfortunately, theoretic CRDT work missed not only these key steps, but also (and more critically) lost sight of the big picture of a co-editing system, thus failing to recognize the nature of CRDT commutativity and real issues, which are discussed in detail in [74,75]. The moral of the story is that it is essential to build working co-editors to know what are the real issues in co-editing and to experimentally validate theoretic solutions. Without the light shed from building co-editors, pure theoretic exploration in the co-editing space could easily be lost in darkness and trapped in pitfalls or illusions, leading to nowhere.

---

[27] https://github.com/atom/teletype
[28] https://github.com/rudi-c/alchemy-book.
[29] https://atom.io/
[30] https://codemirror.net/

Third, *CRDT-based co-editors have all used a central server to support certain aspects of co-editing.* For example, Alchemy Book uses a central server for session management and broadcasting messages among co-editing clients, which is similar to OT-based CoWord [61]; Teletype uses a central server for client-discovery (session management) but allows co-editing clients to be *fully connected* for broadcasting messages without involving the server, which is similar to OT-based REDUCE [56,59]. To our best knowledge, there has been no single example of CRDT-based co-editors that was built without using a client-server architecture. The often-suggested idea that CRDT solutions are specifically designed for peer-to-peer collaborative editing [5,42,46,80,81,82] is tenuous at best, and confounds what CRDT solutions like WOOT and Logoot actually do. We further elaborate this point in Section 4.

Last but not least, by experimenting with Teletype and Alchemy Book prototypes, we can confirm the analytical results (correctness and complexity issues) about tombstone-based (WOOT) and non-tombstone-based (Logoot) CRDT variations, discussed in [75]. In Teletype, we experienced tombstone overhead effects, where the co-editor suffered significant increase of memory consumption and degradation of performance, in both local response and remote replay, as the number of deletions increases during a session. In Alchemy Book, we were able to produce *concurrent-insert-random-interleaving* results when performing concurrent insertions at the same location, and experienced document inconsistencies under numerous scenarios (see concrete examples in [75]). The *concurrent-insert-interleaving* issue was also independently reported by a developer who tried to implement Logoot for text co-editing[31].

*3.2.2 Extending CRDT for Rich Text Co-Editing*

In contrast to most CRDTs for co-editors that are confined to object sequences for supporting plain-text documents, Yjs [38] — a WOOT-like tombstone-based CRDT[32] — is special in extending CRDT to rich-text co-editing.

The Yjs extension for rich-text does not change the core CRDT object sequence and identified-based operations, or make those CRDT-characteristic objects and operations *native* inside any rich-text editor. Instead, the Yjs extension introduces an extra layer of *rich-text-aware* object sequence and operations, which have the knowledge about rich-text features (e.g. **bold**, *italic* attributes), but do not contain tombstones. This extra layer is placed between the Yjs core CRDT solution and the API of a separate rich-text editor (e.g. QuillJS).

Consequently, there exist two object sequences (additional space overheads) outside a rich-text editor: (1) one sequence of rich-text-aware objects (without tombstones); and (2) another sequence of Yjs core CRDT objects (with tombstones but without the knowledge of rich-text). Conversion schemes are devised to resolve data and operation differences among multiple layers, e.g. data objects and operations from the rich-text editor API are converted to/from objects and operations defined on the intermediate rich-text-aware layer (without tombstones), which in turn are converted to/from the CRDT objects (with tombstones) and identifier-based operations in the Yjs core layer. Clearly, main challenges to the Yjs extension lie in the correctness and (space and time) efficiency of those multiple layers and conversion schemes.

---

[31]https://stackoverflow.com/questions/45722742/logoot-crdt-interleaving-of-data-on-concurrent-edits-to-the-same-spot.
[32] https://github.com/y-js/. The core Yjs is based on WOOT but made extensions and changes to WOOT, including a garbage collection scheme for removing tombstones "*after a fixed time period*" [38]. Those changes were intended to reduce the time and space complexity of WOOT, but unfortunately incurred correctness issues (e.g. the garbage collection scheme) within the core Yjs. Detailed elaboration of those issues is beyond the scope of this article.

Again, we observed some conceptual *similarities* between the Yjs extension method and the TA approach (note: we made similar observations for CRDT plain-text co-editors like Teletype and Alchemy Book in Section 3.2.1), as well as their technical *differences* in complexity and efficiency. The success or failure of a co-editor based on a TA-like approach is critically dependent on the *capability*, *correctness* and *complexity* of the core solution (e.g. an OT solution or the Yjs CRDT core solution) and the bridging techniques (e.g. an OT-based collaboration adaptor or the Yjs intermediate layer and conversion schemes) between the core solution and the target editor. The Yjs extension work is still in an early stage of development and its viability remains to be seen. See Section 4.1 of [75] for discussion on some issues in the Yjs CRDT core.

## 3.3 CRDT-based Co-Editing Products

Among existing CRDT-based co-editors, Teletype for Atom is sometimes cited in the CRDT community as an example of industry adoption of CRDT in co-editors. Apart from Teletype, we are not aware of any other industry co-editing product that is based on a CRDT solution.

Why CRDT solutions were rarely adopted in industry co-editing products? Apart from the underlying CRDT correctness and complexity issues (discussed in [75]) and the lack of experimental validation in working co-editors, another obstacle to real world adoption is, in our view, most CRDT solutions for co-editors are confined to inserting and deleting characters or objects in a linear sequence, which are clearly inadequate for supporting real world applications, e.g. rich text formatting, itemization, tables, images, etc. Those real world application features have become *de facto* requirements for collaborative content creation and been commonly supported by OT-based co-editing products, e.g. Google Docs, and Codox Apps. Work on extending CRDT for rich text co-editing, e.g. the Yjs extension, was intended to address this limitation, but still in early stages of development.

## 4 MYTHS AND FACTS ABOUT PEER-TO-PEER CO-EDITING

In spite of the fact that existing co-editors have all been built using a client-server model, CRDT advocators often used "peer-to-peer collaborative editing", or "p2p co-editing" in short, as one primary differentiator between CRDT and OT, and claimed that CRDT is especially suitable for supporting p2p co-editing [5,38,42,46,80,81,82]. One common argument is that CRDT requires no central server, but OT requires a central server (by citing Google Wave/Docs as one evidence, where a central OT-based server is used). Such CRDT p2p arguments have caused quite some confusions in the co-editing community, particularly among new researchers and some practitioners with limited knowledge about OT and co-editing.

In co-editing literature, the term of "p2p co-editing" is rather ambiguous and often conflates multiple factors. So, we tease apart these factors and discuss their relationships to OT and CRDT solutions one-by-one in following subsections.

### 4.1 What Constraints are Imposed on Operation Propagation

To meet the causality-preservation requirement for co-editors [14,55,56], both OT and CRDT solutions generally require editing operations to be executed in their causal-effect orders (based on the *happen-before* relation [25]) at all co-editing sites. This causal ordering can be achieved by adopting any suitable distributed computing technique (based on either client-server or peer-to-peer models), which is orthogonal to OT or CRDT.

However, some CRDT solutions, like WOOT variations, replaced the general causal ordering condition with WOOT-specific execution conditions: an operation can be accepted for execution only if the two neighboring objects (for an insert) or the target object (for a delete) already exists in the internal object sequence. WOOT-specific execution conditions were quoted as a merit for supporting p2p co-editing [42], but such conditions cannot ensure causality-preservation generally required for co-editors [14,55,56], and they are costly to check with the time complexity $O(C_t)$. Other CRDT solutions (e.g. RGA and Logoot variations), as well as OT solutions, are all adopting the general causal ordering condition. Therefore, with the exception of WOOT variations, causally-ordered operation propagation and execution are common requirements for OT and CRDT solutions, and hence not a general differentiating factor between OT and CRDT.

### 4.2 Whether Any Central Server is Required

CRDT solutions do not explicitly require a communication *server*, but all, except WOOT (and its variations), assume the existence of an external causal-order-preserving communication *service*, but left *unspecified* what control information (e.g. timestamping) is required to use this service, and/or whether any server is involved in implementing such a service.

On the other hand, some OT solutions, notably Jupiter [37], NICE [50], and Google Wave and Docs [12,79], explicitly require a *transformation-based* server to do part of the transformation and to broadcast operations (they are *server-based* OT solutions discussed in Section 2.3); SOCT3 and SOCT4 [78] require a server to issue continuous total ordering numbers for labelling operations.

However, most published OT solutions, such as adOPTed [44], GOT[54,56], GOTO[55], SOCT2 [52], TIBOT [27,85], COT [67,68], and POT[85], do not require a central server to do (any part of) the transformation work (they are *server-less* or *distributed* OT solutions discussed in Section 2.3), but require the use of an external causal-order-preserving communication *service*, which is essentially the same as most CRDT solutions. Therefore, whether requiring a server is merely a feature of individual solutions, but not a general factor for differentiating OT and CRDT.

### 4.3 Whether Vector-based or Scalar Timestamping is Used for Control

Yet another factor often cited in connection to p2p co-editing is whether vectors (with one element for each of co-editing sites in a session) or scalars (with a fixed number of variables) are used for timestamping or control in OT or CRDT solutions. Again, both vector-based and scalar-based timestamps and control have been used in OT and CRDT solutions.

Table 1. Summary of p2p co-editing factors and their relationships with OT and CRDT solutions

| P2P Co-Editing Factors | | OT | CRDT |
|---|---|---|---|
| **Requirement for a Central Server** | Yes | Some OT solutions (e.g. Jupiter [37], NICE [50], Google Wave & Docs [12,79]) require a transformation-based server; Some (e.g. SOCT3 & SOCT4 [78]) require a server to issue continuous total ordering numbers. | CRDT solutions (except WOOT) do not explicitly require a communication *server*, but assume the existence of a causal-order-preserving communication *service*. |
| | No | Many OT solutions (e.g. adOPTed [44], GOT[54,56], GOTO[55], SOCT2 [52], TIBOT [27,85], COT [67,68], POT[85]) only require the existence of an external causal-order-preserving communication service. | |
| **Requirement for Operation Propagation or Reception** | Causal Ordering | All OT solutions | All CRDT solutions except WOOT. |
| | Non-causal Ordering | Non | WOOT [42] and variations [4,81] |
| **Timestamping Schemes** | Vector | Some OT solutions (e.g. adOPTed [44], GOT [54,56], GOTO[55], SOCT2 [52], COT [67,68]) use vector-based timestamping. | RGA [46] explicitly requires vector-based timestamping. CRDT solutions (except WOOT) require the existence of an external *causally-ordered* broadcasting service, without specifying what meta control information (or timestamping) needs to be provided to such external service to achieve *causally-ordered* broadcasting. |
| | Scalar | Many OT solutions (e.g. Jupiter [37], NICE [50], Google Wave & Docs [12, 79]; SOCT3 & SOCT4 [78]; TIBOT [27,85], POT[85]) use scalar-based timestamping. | |

For example, the RGA solution uses vector-based timestamps to reduce the time complexity from $O(C_t^3)$ (in WOOT [42]), $O(C_t^2)$ (in WOOTH [4] and WOOTO [81]) to $O(C_t)$, and for garbage collection of tombstones (further reduce $O(C_t)$ to $O(C)$). Logoot [80] and variations [5,82] use variable length (bounded by the object sequence length *C*) object identifiers, and require an external causally-ordered broadcasting service.

On the other hand, some OT solutions, including adOPTed [44], GOT [54, 56], GOTO [55], and COT [67,68], used vector-based timestamps, but other OT solutions, including Jupiter [37], NICE [50], TIBOT [27,85], and SOCT4 [78], Google Wave and Docs [12,79], and POT [85], used scalar-based timestamps. In fact, scalar-based timestamping had been introduced to OT solutions long before the first CRDT (WOOT) came into being. Therefore, neither CRDT nor OT is unique in using vector-based and scalar-based timestamps or control.

In summary (see Table 1), all relevant p2p co-editing factors, namely whether or not requiring a *central server*, *causally ordered communication*, or *vector/scalar timestamping*, are orthogonal to OT and CRDT; each factor is merely features of individual solutions, rather than a general differentiator between OT and CRDT. The notion that CRDT is especially suitable in supporting p2p co-editing is a fallacy.

## 5 CONCLUSIONS

In this work, we have conducted comprehensive and critical reviews of representative OT and CRDT solutions for consistency maintenance in real-time co-editing, and made important research discoveries. We have reported our discoveries in a series of three articles. The main results reported in this article (the third in this series) are summarized below.

In this article, we have focused on examining the main differences between OT and CRDT in building co-editing systems and supporting real world applications. We reviewed the role that building working co-editors has played in OT and CRDT research, respectively. OT research was started and driven mostly by building working co-editors. Researchers have built various OT-based co-editors and used them as research vehicles to identify research issues that are relevant to co-editing, to validate OT solutions under working co-editors, and to explore co-editing system design and implementation issues. The success of OT being adopted in industry products owes in no small part to the research community's persistent efforts in connecting theory with practice and in innovating co-editing system design and implementation solutions. In contrast, CRDT research for co-editing, from its start, has taken predominantly theoretic approaches to identifying research issues, designing and verifying CRDT solutions, but rarely implemented theoretic CRDTs in working co-editors for experimental validation. Such differences have had profound impact in shaping OT and CRDT researches, and in the choice between OT and CRDT in the real world.

By examining OT and CRDT under a general transformation framework, we revealed that published CRDTs had missed several crucial steps in the life-cycle of editing operations from the local generation by a user, all the way to the remote replay on a document visible by another user (reported in the first article [74] of this series). Missing those steps and (more critically) the big picture of a co-editing system in theoretic CRDT work is one root cause for the failure to recognize the non-native commutativity nature of CRDT and the hidden CRDT correctness and complexity issues (reported in the second article [75] of this series). In the current article, we reported our discoveries from examining publically available CRDT-based co-editors in relation to published CRDTs. We found those missing steps in published CRDTs manifested themselves in the source code or documentation of those co-editors, and re-confirmed the overhead issues, algorithmic flaws and abnormalities in representative CRDTs (reported in [75]), as well as the fact that no CRDT is native to any editor. Unfortunately, none of those CRDT co-editors was built by people who published theoretic CRDTs, and various issues manifested from building those co-editors had little impact on theoretic CRDT work. The lesson from the story is that missing the guidance and validation from building co-editors, pure theoretic exploration or paper design in the co-editing space could easily be lost in darkness and trapped in pitfalls or illusions, leading to nowhere.

We have examined and refuted one common misconception in co-editing literature, i.e. CRDT is especially suitable for supporting p2p co-editing as CRDT works without requiring a central server, whereas OT is unsuitable for p2p co-editing as OT requires a central server. The fact is that OT solutions can be either server-based (e.g. Google Wave/Docs OT, often cited by some as the evidence for OT-server-dependence) or server-less (see numerous fully distributed OT solutions elaborated in Section 2.3). Those server-less OT solutions are the same as CRDTs in terms of using no central server. So, it is incorrect to generally differentiate OT and CRDT by whether a server is used in an individual OT or CRDT solution. Furthermore, we have examined other factors related to p2p co-editing, such as whether or not causally ordered communication is required, and whether vector or scalar timestamping is used. We found all those p2p-related factors are orthogonal to OT and CRDT: each factor is merely features of individual OT or CRDT solutions, rather than a general differentiator between OT and CRDT. In a nutshell, the notion that CRDT is especially suitable in supporting p2p co-editing is a fallacy. Up until now, there has been no single p2p co-editor built yet; it remains open whether and how to support p2p co-editing in the future.

In this article, we have also reviewed the evolution of co-editors from research vehicles to real world applications, interactions between the co-editing academic research community and industry, and alternative approaches to building OT-based co-editors in research prototypes, industry products, and open source projects. Over the years, one common co-editing system design approach has emerged, i.e. to separate collaboration capabilities (empowered by OT and/or other suitable techniques) from conventional editing functionalities and encapsulate the two in different functional components (rather than mixing them or making the former a native part of the latter), and provide ways to integrate these components together in a complete co-editing system. This approach represents one major trend in co-editing system design, which is promising to realize the dream of making real-time co-editors commonplace on the Web.

The primary goal of co-editing research is to invent innovative solutions for building useful and useable co-editors. While theoretic work around co-editing algorithms has been fundamental to this endeavor, it is the real world application that provides ultimate validation to co-editing research. We hope discoveries from this work will help clear up common myths and misconceptions surrounding OT and CRDT, inspire fruitful explorations of novel collaboration techniques, and accelerate progress in co-editing technology innovation and real world application.

## 6    ACKNOWLEGMENTS

This research is partially supported by an Academic Research Fund Tier 2 Grant (MOE2015-T21-087) from Ministry of Education Singapore.